\documentclass[sigconf,author]{acmart}
\usepackage{subcaption}
\usepackage{multirow}
\usepackage{multicol}
\usepackage{hyperref} 
\newcommand{\kmy}[1]

\AtBeginDocument{%
  \providecommand\BibTeX{{%
    \normalfont B\kern-0.5em{\scshape i\kern-0.25em b}\kern-0.8em\TeX}}}

\setcopyright{acmcopyright}
\copyrightyear{2023}
\acmYear{2023}
\acmDOI{XXXXXXX.XXXXXXX}

\acmConference[IRS@WSDM '23]{In Proceedings of the Workshop on Interactive Recommender Systems of the 16th ACM International Conference on Web Search and Data Mining}{February 27--March 3,
  2023}{Singapore}
\acmBooktitle{Workshop on Interactive Recommender Systems of the 16th ACM International Conference on Web Search and Data Mining (IRS@WSDM' 23), February 27-March 3, 2023, Singapore}
\acmPrice{15.00}
\acmISBN{978-1-4503-XXXX-X/18/06}
\newcommand{\tabincell}[2]{\begin{tabular}{@{}#1@{}}#2\end{tabular}}

\begin{document}

\title{Improving Recommendation Systems with User Personality Inferred from Product Reviews}



\author{Xinyuan Lu$^{1,2}$ \quad \quad \quad \quad \quad \quad \quad \quad \quad \quad Min-Yen Kan$^2$ \\
$^1$Integrative Sciences and Engineering Programme (ISEP), NUS Graduate School\\
$^2$School of Computing, National University of Singapore, Singapore\\
{\tt luxinyuan@u.nus.edu} \quad \quad \quad \quad \quad \quad {\tt kanmy@comp.nus.edu.sg}\\
}








\renewcommand{\shortauthors}{Lu and Kan}

\begin{abstract}
Personality is a psychological factor that reflects people's preferences, which in turn influences their decision-making. We hypothesize that accurate modeling of users' personalities improves recommendation systems' performance.  However, acquiring such personality profiles is both sensitive and expensive. 
We address this problem by introducing a novel method to automatically extract personality profiles from public product review text. We then design and assess three context-aware recommendation architectures that leverage the profiles to test our hypothesis. 

Experiments on our two newly contributed personality datasets --- {\it Amazon-beauty} and {\it Amazon-music} --- validate our hypothesis, showing performance boosts of 3--28\%.
Our analysis uncovers that varying personality types contribute differently to recommendation performance:
\textit{open} and \textit{extroverted} personalities are most helpful in music recommendation, while a \textit{conscientious} personality is most helpful in beauty product recommendation. 
\end{abstract}





\maketitle

\section{Introduction}
Online recommendation systems are algorithms that help users to find their favorite items. In recommendation systems, the user's profile is important as people with different ages, educational backgrounds exhibit different preferences. Besides static attributes such as gender, the user's psychological factors, especially personality, can be viewed as a user's dynamic profile are also vital in recommendations.

People with similar personalities are more likely to have similar interests and preferences~\cite{DBLP:journals/kbs/YangH19g}. Therefore, accurate modeling of the user's personality plays a vital role in recommendation systems. For example, in movie recommendation, an outgoing person may favour watching comedic movies over romantic ones~\cite{DBLP:journals/kbs/YangH19g}. 
Other studies~\cite{DBLP:conf/ism/MariappanSP12} have shown that in music recommendation, a user's degree of openness strongly determines their preference for energetic music genres. These examples show that personality traits can influence users' preferences.

While we can see that personality traits motivate users' preferences, there are challenges that need to be solved before one can utilize the traits in the recommendation. 
First, collecting personality data is time-consuming. 
The current best practice for collecting personality data requires conducting a user study via an ethically-cleared questionnaire with informed consent. Subsequent training of assessors is also needed. The entire collection process can take months~\cite{DBLP:conf/coling/StajnerY20} and also be an expensive process in terms of effort. 

Second, processing personality data raises sensitivity and privacy concerns.  If handled incorrectly, such data can be misused by users intentionally, resulting in a violation of privacy protection policies and biased performance of the recommendation systems. For example, a scandal emerged when a Facebook app illegally collected 87 million users' personality information to manipulate their voting choices in the U.S. presidential election in March 2018~\cite{DBLP:journals/ijmms/HindsWJ20}. Such risks make the balance between collecting and utilizing users' personality information challenging. This issue has stalled progress in this emerging field of research.

Due to the problem above, the third challenge is a lack of personality-grounded datasets in the existing work. One notable exception is the website \textit{myPersonality}\footnote{\url{https://sites.google.com/michalkosinski.com/mypersonality}}, which contained personality data and the likes of Facebook users. However, in 2018, \textit{myPersonality}'s founders decided to discontinue the project 
as ``complying with various regulations [had] become too burdensome''.  To the best of our knowledge, there are thus few datasets suitable for testing the effect of personality factors in recommendation systems research. 

In this study, we explore methods to overcome these challenges discussed above and contribute to personality-based recommendation research.  We identify a new source for inferring a user's personality traits: user-generated content, specifically e-commerce review texts. Studies show that review texts can reflect a user's personality since individuals manifest their personality through their choice of words~\cite{pennebaker1999linguistic}. For example, when showing their dislikes on the same shampoo, an agreeable person may comment \textit{``I bought this shampoo for my husband. The smell is not good.''}, while a neurotic and aggressive person might comment \textit{``Arrived opened and leaking all over the box. Tried shampoo but didn't help at all. Still so itchy!!!''}. In addition, review text is easy to obtain and made publicly available by users in full disclosure on online commercial websites, which helps to solve both time and privacy issues.

 In our experiments, we explore the possibility of automatically inferring users' personality traits from their review texts and then use this information to help recommendations.  We do this by leveraging an Application Programming Interface (API) to automatically analyze the user's personality. There already exist deployed, production-level APIs for automatic personality detection, such as IBM Personality Insights\footnote{\url{https://cloud.ibm.com/docs/personality-insights}}, Humantic AI\footnote{\url{https://humantic.ai/}}, and Receptiviti\footnote{\url{https://www.receptiviti.com/}} that purport to yield personality profiles. In our work here, we use the Receptiviti API, because it is a widely-validated and widely-used psychology-based language analysis platform for understanding human emotion, personality, motivation, and psychology from language. Receptiviti's API outputs scores for the commonly-used OCEAN personality model: five values, one for each of the five personality aspects of Openness, Conscientiousness, Extroversion, Agreeableness, and Neuroticism (each corresponding to one letter of ``OCEAN''). Finally, this inferred personality is fed as input to a recommendation system, to test whether it can improve recommendation performance. 

To conduct our study, we first construct two new datasets extending from an existing Amazon review dataset, in the beauty and music domains. We first extract the user reviews that are between 30 to 80 words. Afterward, we concatenate all the valid review texts of each user and input their concatenation to the Receptiviti API to output each user's inferred personality scores. 
As a quality check, we evaluate the accuracy of personality detection, by plotting the personality distribution for each dataset.  We observe the users with extremely high/low personality scores and find that these are reliable indicators of personality, and use the such confidently labeled output as ground truth (silver data).

We incorporate these personality scores into the recommendation process and investigate their effect on current neural-based recommendation systems. We observe consistent improvements on the performance of such recommendation systems. 
When we consider different personality groups, we find that \textit{extroversion} and \textit{agreeableness} benefit the recommendation performances across all the domains. 
However, we lack an in-depth understanding of how these personalities affect recommendations and users' behavior. This points to future directions in utilizing other auxiliary information to infer users' personality traits, \textit{e.g.}, users' browsing histories. 

In summary, our contributions are:
\begin{itemize}
    \item We construct two new datasets in the music and beauty domains that combine users' public product reviews alongside automatically inferred personality scores.  This directly addresses the lack of personality-based datasets for recommendation, while avoiding privacy issues by using public data.

    \item We conduct empirical experiments over these datasets, finding that leveraging personality information indeed improves the recommendation performance, from 3\% to 28\%. 
    
    \item We analyze the influence of personality traits in these domains and find the personality traits of extroversion and agreeableness improve the recommendation performance across all domains.
\end{itemize} 

\section{Related Work} 
\label{sec:relatedwork}

The current study investigates how to extract personality traits from texts and how personality traits can be utilized in recommendation systems. Therefore, the review below focuses on the literature that discusses personality detection and personality-based recommendation systems. We first give an introduction of the OCEAN personality models (Section~\ref{sec:ocean}) before reviewing two topics related to our work: personality detection (Section~\ref{sec:personality_detection}) and personality-based recommendation systems (Section~\ref{sec:presonality_recsys}).

\subsection{The OCEAN Model}
\label{sec:ocean}
Personality involves a pattern of behavior that is not likely to change over a short period of time~\cite{allport1961pattern}. It can be detected either explicitly by a questionnaire or implicitly by observing user behaviors~\cite{mccrae1992introduction}. 
The most commonly-used model describing personality traits is the OCEAN model~\cite{mccrae1992introduction}, which we use to model a user's personality traits. The five fundamental personality dimensions defined by OCEAN are: 
\begin{enumerate}
    \item {\textbf{Openness to Experience (O)}}, which describes the breadth and depth of people's life, including the originality and complexity of their experiences. Individuals with high openness tend to be knowledgeable, analytical, and more investigative.
    
    \item {\textbf{Conscientiousness (C)}} This trait involves how individuals control, regulate and direct their impulses. For example, highly conscientious people are usually cautious. 

    \item {\textbf{Extroversion (E)}} Extroversion indicates how much people are in touch with the outside world. Extroverts are more willing to talk to others about their thoughts. 

     \item {\textbf{Agreeableness (A)}} This trait reflects individual differences and social harmony in cooperation. Highly agreeable people are more willing to share tasks than to complete tasks independently.

     \item {\textbf{Neuroticism (N)}} This refers to the tendency of experiencing negative emotions. People with high neuroticism are often in a bad mood, therefore they prefer to respond emotionally.
\end{enumerate}
\subsection{Personality Detection}
\label{sec:personality_detection}

There are two common ways to measure a person's personality traits using a personality model: personality assessment questionnaires and automatic personality detection. 

\subsubsection{\textbf{Personality Assessment Questionnaires}}
Self-reporting personality questionnaires are commonly used to reveal personality differences among individuals. Responses to questions usually take the form of a five-point Likert scale ({\it strongly agree}, {\it agree}, {\it disagree}, and {\it strongly disagree}). Such personality inventories differ with respect to the number and content of their questions. 
Common long questionnaires include the NEO Five-Factor Inventory (60 items)~\cite{ALUJA2005591}, NEO-Personality-Inventory Revised (240 items)~\cite{costa2008revised}, and the Big-Five Inventory (BFI, 44 items)~\cite{rammstedt2007measuring}. 
Practitioners prefer using shorter instruments, such as the BFI-10 and Ten-Item Personality Inventory (TIPI)~\cite{gosling2003very,topolewska2014short}, as they are time-saving and easier to fill out.  

However, using questionnaires for self-report has two major drawbacks. First, questions that assess personality are often quite subjective such as ``Do you easily get nervous?''. Answers for such questions are easily affected by a self-bias~\cite{PEDREGON2012213} or reference-group effects~\cite{doi:10.1177/0956797616678187}. For example, an introverted engineer might think he is an extrovert if he/she is working with a group of individuals that may be more introverted. Consequently, the results of questionnaires are often hard to reproduce. Second, assessing a personality by questionnaires can be inconvenient, as the subjects are necessary to participate in the studies. 

\subsubsection{\textbf{Automatic Personality Detection}} 
To make personality detection more convenient and reproducible, practitioners prefer automated personality detection, which infers a personality type based on user data. 
Such methods are less accurate than personality questionnaires --- as it relies on the input user data manifesting personality traits --- but has the advantage of not requiring inputs to be answers to questionnaires.  For example, social media posts that exhibit opinions and viewpoints are a prime source of text data useful for personality detection. Individuals have different language use behaviors that reveal personality traits~\cite{hirsh2009personality}. Automatic, text-based personality detection infer users' personalities by analyzing users' word choice (lexical selection) and sentence structure (grammatical selection). Such technology has been sufficiently proven, making them commonplace and deployed at scale in production, and available as a service through cloud-based application APIs.

We study whether knowing users' personality information can lead to better recommendations, and also, how users' personality information be best modeled in recommendation systems to improve performance. While large-scale recommendation datasets exist, they universally lack users' personality information. It is infeasible to ask to obtain this information via questionnaires since the identity of users is usually confidential. Therefore, we utilize automatic personality detection to infer personality from product reviews written by users.  Product reviews are ideal: they are widely available on online commercial websites, they often demonstrate personality traits, and they are public (the texts are meant to be read by others).  Hence, they can serve as a good source for automatically detecting personality in an economic but accurate way. 

\subsection{Personality-based Recommendation Systems}
\label{sec:presonality_recsys}

Since personality traits are characteristics that do not change sharply over time and do not depend on a certain context or stimulus, they are more easily used to create personalized recommendation systems. Earlier work by Winoto and Tang~\cite{DBLP:journals/eswa/WinotoT10}~\cite{ayata2018emotion} focused on extending Matrix Factorization by adding a personality latent factor. Their model used implicit feedback data, such as user--item interactions, beyond just ratings. 
They only considered the OCEAN scores as one attribute, so the effects that are attributable just to personality are not clear. Besides, personality traits have been used to determine a neighborhood of similar users by calculating personality similarity. Thus, for example, Asabere \textit{et al.}~\cite{DBLP:journals/taffco/AsabereAM18} proposed a recommendation system for attendees of conferences that integrates personality traits and social bonds of attendees. 

In their work, user similarity was equated as personality similarity, calculated by Pearson’s correlation between two users' OCEAN scores. They demonstrated that the system accuracy improves with a larger user base, due to the higher likelihood of finding other users with similar personalities (high correlation).

Researchers have also associated user personality scores with items. Yang and Huang~\cite{DBLP:journals/kbs/YangH19g} attributed items (here, computer games) with a personality that is an average of its users. This latent representation can then be used to recommend items to users with a similar personality as that of the other users of that item.  This may make sense when certain items are used primarily by certain personality types (as in computer games) but are less compelling for items that may be used by many personality types.
Lastly, in social media recommendations, Wu \textit{et al.}~\cite{DBLP:journals/umuai/WuCZ18} proposed an approach for recommending interest groups by integrating personality. The personality-based similarity was defined as the Euclidean distance between two users’ personality scores. However, it combines the personality signal linearly in the recommendation process, which we feel may be limiting.  

In summary, 
compared with other context attributes (e.g., purchase history), personality information helps to capture the users' potential interests rather than recommending a similar purchased item. However, the previous works  used the OCEAN personality as a linear similar score which lacks the capability of capturing more nuanced information latent in personality scores.  Different from the methods above, we propose two novel ways of adding personality features into the recommendation system: 1) taking the most salient personality trait as a learnable vector and 2) calculating a user's personality embedding as a weighted sum of a user's OCEAN personality features, which is a learnable embedding within the recommendation system.

\section{Dataset Construction}
\label{sec:dataset}
We construct two new datasets, as extensions of the existing, well-known Amazon review dataset. 
We first automatically infer users' personality traits from users' review texts as review texts can reflect the personality through  word usage. They are also publicly-available text on online commercial websites, allowing researchers to have legal access to textual data where experimentation can be replicated. 
Based upon the parent Amazon review dataset, we construct two new domain-specific datasets: an \textit{Amazon-beauty} and an \textit{Amazon-music} dataset.  These contain Amazon reviews of products in the beauty and music domains, alongside their posting users' inferred personality scores.

\subsection{Data Source}
\label{sec:data source}

The Amazon dataset\footnote{\url{https://nijianmo.github.io/amazon/index.html}}~\cite{DBLP:conf/emnlp/NiLM19} is widely used for training and evaluating recommendation systems. It contains a large number of item descriptions, ratings, and product reviews collected from the Amazon online commercial website. The Amazon dataset is divided according to the domain. In our study, we choose two domains: \textit{beauty} and \textit{music}. 
We construct datasets separately for these two domains since we want to study whether personality has different influences on users' behaviours for different domains. Studies have shown that people with different personalities prefer different kinds of music~\cite{DBLP:conf/ism/MariappanSP12}. For example, people with a high degree of openness like to listen to rock music, while neurotic people like jazz. Therefore, we choose \textit{music} as one of the domains to be studied. In order to study the role of personality in different domains, we randomly select \textit{beauty} for comparison. Table~\ref{tab:amazon good saample} shows a sample of the original Amazon dataset, which contains the user (\textit{reviewerID}, \textit{reviewerName}), the product's Amazon Standard Identification Number (\textit{asin}), the review text for the product (\textit{reviewText}), and the overall rating given to the product (\textit{overall}). 

\begin{table}[!ht]
\centering
\resizebox{0.47\textwidth}{!}{\begin{tabular}{c|c}\hline
\textbf{reviewerID} & A2SUAM1J3GNN38 \\ \hline
\textbf{asin} & 0000013714 \\ \hline
\textbf{reviewerName} & J.McDonald \\ \hline
\textbf{vote} & 5 \\ \hline
\textbf{style} & Format:Hardcover \\ \hline
\textbf{reviewText} & \tabincell{l}{I bought this for my husband who plays the piano. He is \\ having a wonderful time playing these old hymns. The \\ music is at times hard to read because we think the book \\was published for singing from more than playing form. \\Greate purchase though!}\\ \hline
\textbf{overall} & 5.0 \\ \hline
\end{tabular}
}
\vspace{+3mm}
\caption{An example of Receptiviti score for a specific, anonymized user.}
\label{tab:amazon good saample}
\end{table}

\subsection{Dataset Construction}
\label{sec: dataset construction}

Since we do not know the personality for each user in the Amazon dataset, we need to infer them. We first retrieve each user's review texts and then use the Receptiviti API\footnote{\url{https://www.receptiviti.com/}}, a computational language psychology platform for understanding human behavior, to infer a personality. The API can take a long piece of human-written text (more than 300 words), and output a faceted personality score with 35 factors, including OCEAN scores.  


For each user that wrote reviews in either of the two domains, we collect all his/her review texts and concatenate them together into a single document. Afterward, we send the concatenated text to Receptiviti to infer a personality. We select the personality scores corresponding to the five-dimensional personality traits defined in the OCEAN model~\cite{mccrae1992introduction} (Table~\ref{tab:receptiviti sample}).  Each personality score is normalized to a range from 1 to 100. The higher the score, the more overt the personality trait. Note that each of the five OCEAN scores is independent of the other.
\begin{table}[!ht]
\centering
\resizebox{0.38\textwidth}{!}{\begin{tabular}{|c|c|c|c|c|c|}\hline
\textbf{User ID} & \textbf{{AGR}} & \textbf{{CON}}& \textbf{{NEU}} & \textbf{{EXT}} & \textbf{{OPEN}} \\ \hline
A2GBIFL43U1LKJ & 54.05 & 34.87 & 25.96 & 54.39 & 42.71 \\ \hline
\end{tabular}
}
\vspace{+3mm}
\caption{An example of Receptiviti score for a specific, anonymized user.}
\label{tab:receptiviti sample}
\end{table}

To improve the personality prediction process, we only analyze the personality traits for \textit{active} users who bought many products and wrote a sufficient number of product reviews. To be specific, we select users that 1) wrote product reviews for at least 10 different items they purchased, and where 2) 
each product review contains between 30 to 80 words. 
Table~\ref{tab:dataset} shows the statistics after the filtration. 
For example, using these criteria, 1,791 \textit{active} users are selected for the \textit{Amazon-music} dataset. Each user in the {\it Amazon-music} dataset has an average of 990.48 review words over all of his/her reviews, averaging 51.01 words for each review. 



\subsection{Dataset Statistics}

Aside from our constructed \textit{Amazon-beauty} and \textit{Amazon-music dataset}, we also include an existing dataset \textit{Personality 2018} in our study. Personality 2018\footnote{\url{https://grouplens.org/datasets/personality-2018/}}~\cite{DBLP:journals/isf/NguyenHTK18} is a version of the MovieLens dataset that includes each user's personality information obtained through questionnaires. It contains 21,776 movies, 339,000 ratings, and 678 users with the OCEAN personality questionnaire scores from 1 to 7. This dataset is included to study the difference between questionnaire-based personality trait scores with our review-based automatic personality trait detection scores.  

Table~\ref{tab:dataset} shows the final statistics of the datasets used in our study. We can observe that the \textit{Amazon-beauty} / \textit{Amazon-music} dataset has the largest / smallest percentage of interactions. The \textit{Personality2018} dataset contains the largest number of items and the smallest number of users.  
We can see that these datasets differ in domains, number of users, items, and interactions, which facilitates the study of personality-based recommendation across a wide spectrum of settings. 


\begin{table}[ht]
\centering
\resizebox{0.5\textwidth}{!}{
\begin{tabular}{c|c|c|c}\hline

{\bf Dataset} & \textbf{Amazon-beauty} & \textbf{Amazon-music} & \textbf{Personality'18} \\\hline
\# of items & 85 & 8,895 & 21,776\\\hline
\# of users & 991 & 1,791 & 678\\ \hline
\# of ratings & 5,269 & 28,399 & 339,000 \\ \hline
\# of interactions & 6.26\% & 0.18\% & 2.30\%\\ \hline \hline
\tabincell{c}{Avg. words/user} & 990.48 & 466.43 &  - \\ \hline
\tabincell{c}{Avg. words/review} & 51.01 & 51.18 &  - \\ \hline

\end{tabular}
}
\vspace{+3mm}
\caption{Statistics of the three datasets used in our study.}
\label{tab:dataset}
\end{table}

\section{Methods}
\label{sec:method}

Based on our constructed dataset, we conduct experiments to study whether the recommendation system can benefit from incorporating personality traits. We choose the Neural Collaborative Filtering (NCF)~\cite{DBLP:conf/www/HeLZNHC17} as the foundation model of our study because it is the fundamental neural-based model for the recommendation. Specifically, we design a personality-enhanced version of NCF~\cite{DBLP:conf/www/HeLZNHC17} to compare with the vanilla NCF, alongside several other baselines.

\subsection{Neural Collaborative Filtering (NCF)} 
NCF~\cite{DBLP:conf/www/HeLZNHC17} is the first deep-learning-based recommendation algorithm. Different from traditional collaborative filtering algorithms, the model encodes the user and item into latent vectors and then projects them through a Multi-layer Perceptron (MLP) to predict a probability score, representing the probability that a user would buy a target item. In our implementation, we use a 4-layer MLP and a 16-dimensional user and item embedding. 

\subsection{Personality-enhanced NCF} 
\begin{figure}[hbt!]
  \centering
  \includegraphics[width=9.2cm]{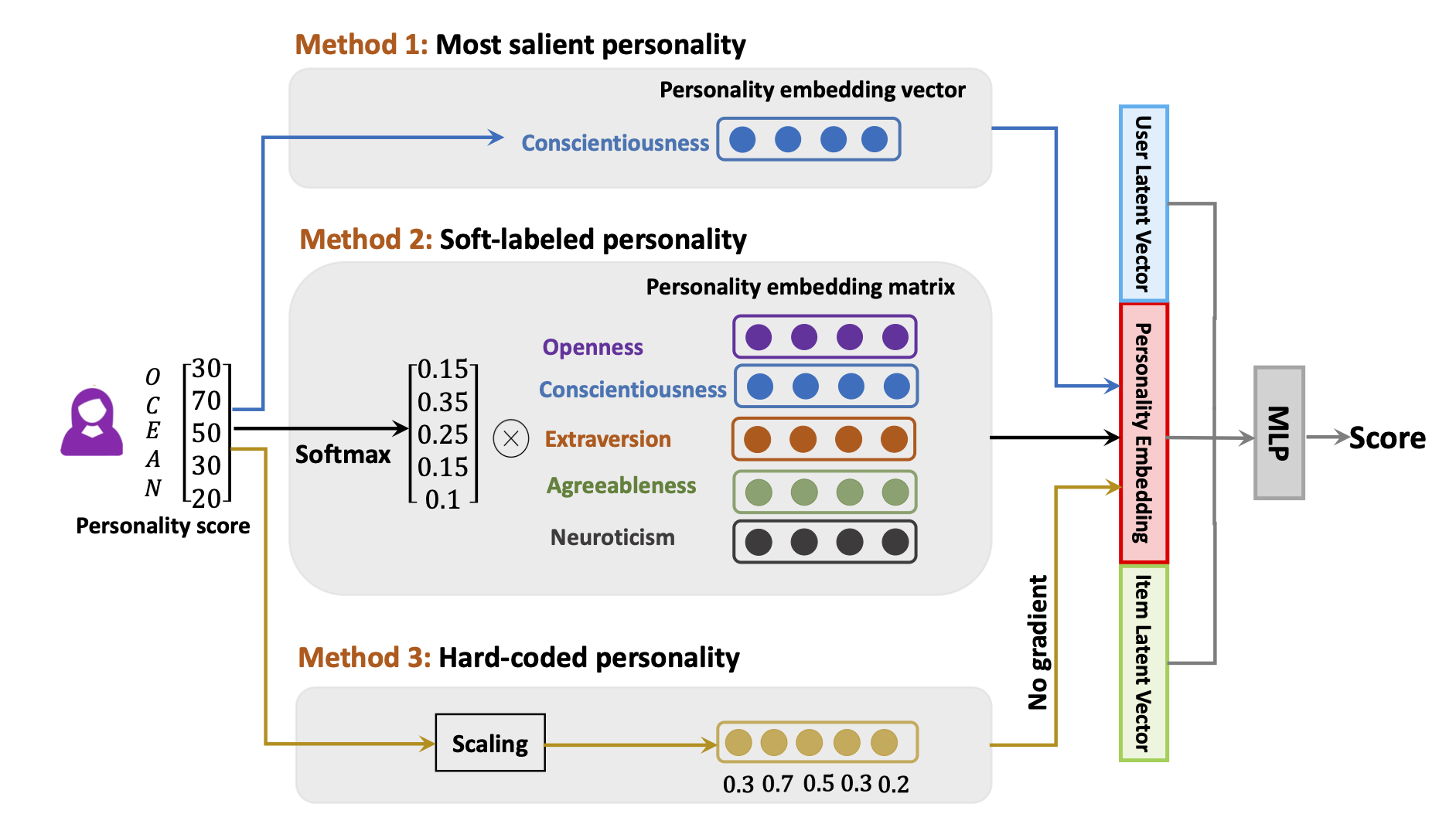}
  \caption{The overall structure of our model. In this example, the user's OCEAN score is \{30,70,50,30,20\}. The\textbf{NCF + Most salient personality} selects the personality with the highest score, i.e., \textit{conscientiousness} as the personality embedding vector. \textbf{NCF + Soft-labeled personality} takes all five OCEAN scores as a personality embedding matrix.  
  \textbf{NCF + Hard-coded personality} predetermines and fixes the personality vector as \{0.3,0.7,0.5,0.3,0.2\}
  \label{fig:model}}
\end{figure}

We then propose three different ways to incorporate the personality information into the NCF model, as shown in Fig.~\ref{fig:model}. We first design \textit{NCF+Most salient Personality} model by adding the most salient personality trait as input into NCF. We also design \textit{NCF + Soft-labeled Personality} and \textit{NCF + Hard-coded Personality} to incorporate all the five personality traits of OCEAN. The difference between the two latter versions is that the personality vector in \textit{NCF + Soft-labeled Personality} is learnable, while in \textit{NCF + Hard-coded Personality}, the vector is predetermined and fixed. 

1. \textbf{NCF + Most Salient Personality.} In this model, we introduce a 4-dimensional \textit{personality vector} for each of the five types of personalities, which are learned during training. We treat the most salient personality as the user's personality label and concatenate the corresponding personality vector with the user's latent vector. 

2. \textbf{NCF + Soft-labeled Personality.} In this model, we make full use of all five personality trait scores. We first apply a \textit{Softmax} function to map the personality scores into a probability distribution of personality. Afterward, the probability distribution is used as the weight to calculate the weighted sum of the five personality vectors. The output vector is then concatenated with the user's latent vector as the input of the MLP. 

3. \textbf{NCF + Hard-coded Personality.} This model also considers all the user's five personality traits information. However, instead of introducing learnable personality vectors, we directly scale each personality score to sum to a unit value (here, 100) to get a hard-coded 5-dimensional vector to represent the user's personality information. This vector is concatenated with the user's latent vector, but is fixed during training.

\section{Experiments}
\label{sec:experiment}
We evaluate our proposed method on our three datasets by answering the following four research questions. We first evaluate whether we can accurately detect personality from texts (RQ1, Section~\ref{sec:RQ1}). Afterward, we analyze the distribution of personality in review texts (RQ2, Section~\ref{sec:RQ2}). Then, we explore whether adding personality information can improve recommendation performance (RQ3, Section~\ref{sec:RQ3}). Finally, we analyze the influence of personality information on different domains (RQ4, Section~\ref{sec:RQ4}). 

\subsection{Can we accurately detect personality from texts? (RQ1)}
\label{sec:RQ1}

To evaluate whether we can accurately detect personality traits from texts, we analyze the personality scores inferred by the Receptiviti API for each user. Since there are over 2,500 users in total in our two constructed datasets, it is time-consuming to manually evaluate them all. As a compromise, we choose to manually examine the users that receive extremely high scores for certain personality traits. We believe those examples are more easily evaluated by humans. Specifically, for each personality trait, we select the users that receive the top 10 highest scores on this type. We analyze both the \textit{Amazon-beauty} and the \textit{Amazon-music} datasets, resulting in a total of 100 samples. 
These samples are evaluated by two graduate students. Both were trained with a detailed explanation of the OCEAN personality model. We ask them to choose whether the sampled review texts  accurately match their inferred personality, choosing between three options of {\it yes}, {\it no}, or {\it not sure}. We then calculate the accuracy of the samples and the inter-annotator agreement between the two annotators using Cohen's Kappa~\cite{Cohen1968WeightedKN}. 
We find that the inferred personality matches with the review text in 81\% of the \textit{Amazon-beauty} samples, and 79\% of the samples from \textit{Amazon-music}. The average Cohen's Kappa is 0.70. We take this to indicate that the Receptiviti API can indeed infer users' personality traits from review texts with generally high accuracy. 

Table~\ref{tab:data sample} shows examples of review texts with their inferred personality scores. We observe that people with different personalities have different language habits. For example, extroverts  tend to use the words ``love'' and exclamation marks because they are characterized by a strong tendency to express their affection. People who are agreeable are usually bought items for other people, \textit{e.g.,} ``my kids'' and ``my wife'', perhaps due to their inclusiveness. 
Conscientious people usually talk about their own experience and feelings before recommending the items to others, \textit{e.g.,} ``I have had this shower gel once before'' or ``Don't just take my word for it''. This is perhaps because they are usually cautious. 
 
\begin{table*}[!hbt]
 \centering
\begin{tabular} {|c|c|l|} \hline
 Personality label  & \tabincell{c} {Personality\\ Score}  & Review Texts \\ \hline
Openness & 63.07 &  \tabincell{l} {Near perfect exfoliating gloves my only complaint is a matter of preference rather than product\\ defect.\\  I prefer the harder surface area to use on round areas of the body or potentially harder like the feet,\\ elbows, etc.} \\ \hline
 Openness & 62.62 & \tabincell{l} {Azur is always my favorite in the Thymes collection because of its clean, fresh scent. \\I like that my skin feels moisturized when using this product in the shower. }\\ \hline
Conscientiousness & 75.38 & \tabincell{l} {I have had this shower gel once before, and it's amazing. Hard to find, too. \\One of The Body Shop's best scents, and it's usually only available seasonally!\\ wish they sold it in bigger bottles, but I was happy to find it.} \\ \hline
 Conscientiousness & 71.02 & \tabincell{l}{Don't just take my word for it, you must try it. \\A dear friend got me this from Italy 12 years ago and has been using it since, \\very hard to find it in the US.\\ This shower cream will transform your shower experience.} \\ \hline
Extroversion & 75.06 & Love this shampoo! Recommended by a friend! The color really lasts!!!\\ \hline
 Extroversion &72.90& \tabincell{l} {Looked all over to find where to purchase this product and we are very happy to \\ be able to finally find it. \\ The PRELL Conditioner is by far the best you can buy. We love it!!} \\ \hline
 Agreeableness & 80.06 & Great product - my wife loves it \\ \hline
 Agreeableness & 78.18 & \tabincell{l} {Great deal and leaves my kids smelling awesome!\\ I bought a box of them years ago and we still have some left!!! \\ Great deal and leaves my kids smelling awesome!} \\ \hline
 Neuroticism & 67.81 & \tabincell{l} {Too expensive for such poor quality. \\There was no improvement and I am starting to think my scalp is worse off than it was before \\ I started using this product.\\ I do agree with other reviews that it feels watered.} \\ \hline
 Neuroticism & 62.28 & \tabincell{l} {Nope. It smells like artificial bananas, and this smell does linger. \\It's pure liquid, there is no thickness to it at all, it's like pouring banana water on your head \\that lathers. \\It does not help with an itchy scalp either.} \\
\hline
\end{tabular}
\vspace{+3mm}
 \caption{The data sample  of extreme personality cases to the annotators. Each data sample contains the user's personality labels, personality scores, and review texts.}
 \label{tab:data sample}
\end{table*}

\subsection{What is the distribution of users' personalities? (RQ2)}
\label{sec:RQ2}
\begin{figure*}[hbt!]
  \centering
  \includegraphics[width=18cm]{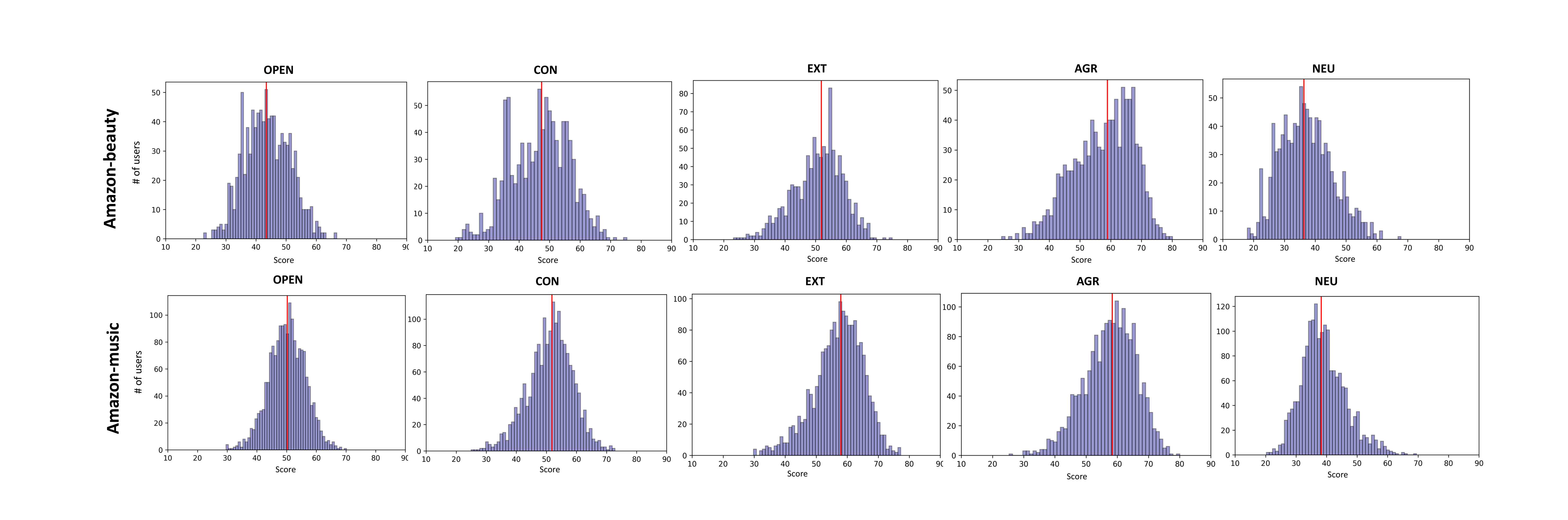}
  \caption{Distribution of personality traits in \textit{Amazon-beauty} and \textit{Amazon-music} datasets. The x-axis represents the score for each trait; the y-axis represents the number of users. The red line represents the median for each trait.}
  \label{fig:personalitydistribution}
\end{figure*}

We further analyze the personality distribution for all users by plotting the score histograms for each personality trait in the \textit{Amazon-beauty} dataset and the \textit{Amazon-music} dataset in Fig.~\ref{fig:personalitydistribution}.




We observe a similar trend in both domains: agreeable people have the highest median score, and neurotic people have the lowest median score. A possible reason is that neurotic people are more introverted and are less likely to publish their opinions publicly, while agreeable people are more willing to share their thoughts. Another observation is that the personalities of each dataset are generally bell-curved. This indicates that each personality trait is normally distributed.

We also examine the difference in personality distributions between the two domains. In the \textit{Amazon-music} dataset, the average scores of extroversion and openness are higher than those in the \textit{Amazon-beauty} dataset. This indicates that the personality characteristics of extroverts and open people are more obvious in the music domain than in the beauty domain. 

From the above figures, we draw the following conclusions. First, the personality traits of users are not evenly distributed. There are more instances of people with certain personality traits (\textit{e.g.,} agreeableness) than others (\textit{e.g.,} neuroticism). A possible reason is that people with certain personalities are more willing to write product reviews. 
2) The distributions for the two domains are generally the same, with higher agreeable scores and lower neurotic scores. However, there is a slight difference. For example, the scores of extroverts in music are generally higher than that in the beauty domain. This could be explained by the possibility that people who are passionate about music may be more emotional. 
\subsection{Does incorporating personality improve recommendation performance? (RQ3)}
\label{sec:RQ3}
Next, we want to explore whether adding the induced user personality benefits the recommendation quality. To this end, we compare the personality-enhanced NCF with the following two baseline models that do not utilize personality information. 

1. \textbf{NCF with random personality (NCF + Random).} We randomly assign each user with a random personality label, regardless of his/her original, inferred personality scores.  

2. \textbf{NCF with same personality (NCF + Same).} We assign each user to a single personality trait. To be specific, we assume every user is ``open'' and assign the corresponding personality vector to NCF. Although the personality vector does not provide any additional signal to the model in this case, it can serve as a place-holder to keep the network structure identical to the personality-enhanced model, resulting in a fair comparison. 

\textbf{Evaluation Metrics.}
We use two metrics to measure the performance of our proposed recommendation models: \textit{Hit Rate (HR) @ K} and \textit{Normalized Discounted Cumulative Gain (NDCG) @ K (K = 3, 5, 10)}. 
Larger HR and NDCG demonstrate better accuracy.  

\textbf{Experiment Results.}
Table~\ref {tab:experiment} shows the experimental results in the \textit{Amazon-beauty} and the \textit{Amazon-music}, and \textit{Personality2018} datasets, respectively. In \textit{Amazon-beauty} and \textit{Amazon-music}, we find that the three personality-enhanced NCF models outperform the two baseline models, in terms of both NDCG and HR. Especially, the first three rows show that the NCF with the most salient personality label outperforms NCF with the same or random personality label. This indicates that adding personality information into NCF improves recommendation performance. From the last three rows, we further find that NCF + Soft-labeled/Hard-coded outperforms NCF + Most salient personality in terms of NDCG. This shows that utilizing all five personality traits is better than using the most salient personality trait in NCF.

In the \textit{Personality 2018} dataset, the trend in the \textit{Amazon-beauty} and \textit{Amazon-music} also holds for it. For example, the NCF + Soft-labeled model outperforms the other models, showing that adding personality information improves performance. However, the improvement in the \textit{Personality 2018} is less obvious than that in the \textit{Amazon-beauty} dataset. We hypothesize the reason might be due to the difference in the sizes of the datasets. Since \textit{Amazon-beauty} is a small dataset, adding personality information may better help to address the data sparsity problem, therefore exhibiting a better performance gain. 

\begin{table*}[!htb]
\resizebox{\textwidth}{!}{\begin{tabular}{c|cllcll|cllcll|cllcll}
\hline
Algorithms       & \multicolumn{6}{c|}{\textit{\textbf{Amazon-beauty}}}                                                                                                                                                         & \multicolumn{6}{c|}{\textit{\textbf{Amazon-music}}}                                                                                                                                                      & \multicolumn{6}{c}{\textit{\textbf{Personality2018}}}                                                                                                                                                      \\ 
\hline
Rating           & \multicolumn{1}{c|}{\textbf{H@3}}   & \multicolumn{1}{l|}{\textbf{H@5}}   & \multicolumn{1}{l|}{\textbf{H@10}}  & \multicolumn{1}{c|}{\textbf{N@3}}   & \multicolumn{1}{l|}{\textbf{N@5}}   & \textbf{N@10}  & \multicolumn{1}{c|}{\textbf{H@3}} & \multicolumn{1}{l|}{\textbf{H@5}} & \multicolumn{1}{l|}{\textbf{H@10}}  & \multicolumn{1}{c|}{\textbf{N@3}}   & \multicolumn{1}{l|}{\textbf{N@5}}   & \textbf{N@10}  & \multicolumn{1}{c|}{\textbf{H@3}}   & \multicolumn{1}{l|}{\textbf{H@5}}   & \multicolumn{1}{l|}{\textbf{H@10}}  & \multicolumn{1}{c|}{\textbf{N@3}}   & \multicolumn{1}{l|}{\textbf{N@5}}   & \textbf{N@10}  \\ \hline
NCF+Random       & \multicolumn{1}{c|}{0.923}          & \multicolumn{1}{l|}{0.965}          & \multicolumn{1}{l|}{0.975}          & \multicolumn{1}{c|}{0.675}          & \multicolumn{1}{l|}{0.605}          & 0.660          & \multicolumn{1}{c|}{0.159}        & \multicolumn{1}{l|}{0.224}        & \multicolumn{1}{l|}{0.339}          & \multicolumn{1}{c|}{0.117}          & \multicolumn{1}{l|}{0.143}          & 0.171          & \multicolumn{1}{c|}{0.510}          & \multicolumn{1}{l|}{0.628}          & \multicolumn{1}{l|}{0.777}          & \multicolumn{1}{c|}{0.406}          & \multicolumn{1}{l|}{0.454}          & 0.504          \\ \hline
NCF+Same         & \multicolumn{1}{c|}{0.918}          & \multicolumn{1}{l|}{0.967}          & \multicolumn{1}{l|}{0.975}          & \multicolumn{1}{c|}{0.683}          & \multicolumn{1}{l|}{0.630}          & 0.662          & \multicolumn{1}{c|}{0.160}        & \multicolumn{1}{l|}{0.224}        & \multicolumn{1}{l|}{0.340}          & \multicolumn{1}{c|}{0.122}          & \multicolumn{1}{l|}{0.149}          & 0.167          & \multicolumn{1}{c|}{0.511}          & \multicolumn{1}{l|}{0.622}          & \multicolumn{1}{l|}{0.777}          & \multicolumn{1}{c|}{0.403}          & \multicolumn{1}{l|}{0.454}          & 0.502          \\ \hline \hline
NCF+Most-Salient & \multicolumn{1}{c|}{0.939}          & \multicolumn{1}{l|}{\textbf{0.969}} & \multicolumn{1}{l|}{\textbf{0.977}} & \multicolumn{1}{c|}{0.714}          & \multicolumn{1}{l|}{0.676}          & 0.707          & \multicolumn{1}{c|}{0.156}        & \multicolumn{1}{l|}{0.226}        & \multicolumn{1}{l|}{0.343}          & \multicolumn{1}{c|}{\textbf{0.164}}          & \multicolumn{1}{l|}{0.145}          & 0.174          & \multicolumn{1}{c|}{0.516}          & \multicolumn{1}{l|}{0.631}          & \multicolumn{1}{l|}{0.795}          & \multicolumn{1}{c|}{0.415}          & \multicolumn{1}{l|}{0.463}          & 0.511          \\ \hline
NCF+Soft-labeled & \multicolumn{1}{c|}{0.936}          & \multicolumn{1}{l|}{0.965}          & \multicolumn{1}{l|}{0.973}          & \multicolumn{1}{c|}{0.810}          & \multicolumn{1}{l|}{\textbf{0.867}} & 0.831          & \multicolumn{1}{c|}{0.156}        & \multicolumn{1}{l|}{0.225}        & \multicolumn{1}{l|}{\textbf{0.348}} & \multicolumn{1}{c|}{0.113}          & \multicolumn{1}{l|}{0.141}          & 0.175          & \multicolumn{1}{c|}{\textbf{0.528}} & \multicolumn{1}{l|}{\textbf{0.656}} & \multicolumn{1}{l|}{\textbf{0.805}} & \multicolumn{1}{c|}{\textbf{0.421}} & \multicolumn{1}{l|}{\textbf{0.471}} & \textbf{0.511} \\ \hline
NCF+Hard-Coded   & \multicolumn{1}{c|}{\textbf{0.948}} & \multicolumn{1}{l|}{0.961}          & \multicolumn{1}{l|}{0.977}          & \multicolumn{1}{c|}{\textbf{0.849}} & \multicolumn{1}{l|}{0.826}          & \textbf{0.848} & \multicolumn{1}{c|}{0.175}        & \multicolumn{1}{l|}{0.232}        & \multicolumn{1}{l|}{0.345} & \multicolumn{1}{c|}{0.147} & \multicolumn{1}{l|}{\textbf{0.160}} & \textbf{0.189} & \multicolumn{1}{c|}{0.503}          & \multicolumn{1}{l|}{0.622}          & \multicolumn{1}{l|}{0.758}          & \multicolumn{1}{c|}{0.398}          & \multicolumn{1}{l|}{0.447}          & 0.498          \\ \hline
\end{tabular}
}
\vspace*{+3mm}
\caption{Hit Rate(H) and NDCG(N) @K in the \textit{Amazon-beauty}, \textit{Amazon-music}, and  \textit{Personality 2018} datasets. The best performance is bolded.}
\label{tab:experiment}
\vspace{-8mm}
\end{table*}

\subsection{How does personality information improve the performance of recommendation system? (RQ4)}
\label{sec:RQ4}

To gain a better understanding of the improvement brought by incorporating personality, we separately evaluate the HR and NDCG for the five personality traits, as shown in Table~\ref{tab:analysis} . ``+'' represents the NCF+Soft-labeled model (with personality information), and ``-'' represents the NCF+Same model (without personality information). We make two major observations. 

First, the improvement brought by adding personality is prominent for the \textit{Amazon-beauty} dataset, over all five personality traits. In particular, the trait of conscientiousness (CON) has the highest gain in terms of both HR (+21\%) and NDCG (+57\%). However, in the \textit{Amazon-music} dataset, \textit{openness} (+27\%), \textit{agreeableness} (+10\%), \textit{extroversion} (+5\%) improve while \textit{neuroticism} (–18\%) and \textit{conscientiousness} (–12\%) decreases. 

Second, for the \textit{Personality2018}  dataset, the improvement brought by adding personality is not obvious: only \textit{conscientiousness}, \textit{extroversion}, and \textit{agreeableness} have shown minor performance gain. From the above breakdown analysis, we find that adding personality information can benefit certain personality traits better than others. However, the personality trait that improves the most differs greatly across the three datasets. This indicates that although improvements are observed in terms of empirical results, the mechanism of how personality influences the recommendation still deserves more in-depth investigation.

\begin{table}[htb!]
\resizebox{0.5\textwidth}{!}{\begin{tabular}{c|c|cc|cc|cc}
\hline
&  & \multicolumn{2}{c|}{\textit{\textbf{Amazon-beauty}}} & \multicolumn{2}{c|}{\textit{\textbf{Amazon-music}}}  & \multicolumn{2}{c}{\textit{\textbf{Personality2018}}} \\ 
Trait                 &   & \multicolumn{1}{c}{HR}             & NDCG           & \multicolumn{1}{c}{HR}             & NDCG           & \multicolumn{1}{c}{HR}             & NDCG           \\ \hline
\multirow{2}{*}{OPEN} & + & \multicolumn{1}{c|}{\textbf{0.833}} & \textbf{0.729} & \multicolumn{1}{c|}{\textbf{0.330}} & \textbf{0.205} & \multicolumn{1}{c|}{\textbf{0.535}} & 0.420          \\ 
                      & - & \multicolumn{1}{c|}{0.750}          & 0.545          & \multicolumn{1}{c|}{0.313}          & 0.161          & \multicolumn{1}{c|}{0.547}          & \textbf{0.422} \\ \hline
\multirow{2}{*}{CON}  & + & \multicolumn{1}{c|}{\textbf{0.883}} & \textbf{0.769} & \multicolumn{1}{c|}{0.228}          & 0.132          & \multicolumn{1}{c|}{\textbf{0.475}} & 0.358          \\ 
                      & - & \multicolumn{1}{c|}{0.727}          & 0.490          & \multicolumn{1}{c|}{\textbf{0.279}} & \textbf{0.150} & \multicolumn{1}{c|}{0.441}          & \textbf{0.361} \\ \hline
\multirow{2}{*}{EXT}  & + & \multicolumn{1}{c|}{\textbf{0.970}} & \textbf{0.882} & \multicolumn{1}{c|}{\textbf{0.319}} & \textbf{0.181} & \multicolumn{1}{c|}{\textbf{0.611}} & \textbf{0.412} \\ 
                      & - & \multicolumn{1}{c|}{0.872}          & 0.600          & \multicolumn{1}{c|}{0.317}          & 0.169          & \multicolumn{1}{c|}{0.556}          & 0.411          \\ \hline
\multirow{2}{*}{AGR}  & + & \multicolumn{1}{c|}{\textbf{0.968}} & \textbf{0.878} & \multicolumn{1}{c|}{\textbf{0.332}} & \textbf{0.198} & \multicolumn{1}{c|}{\textbf{0.621}} & \textbf{0.512} \\ 
                      & - & \multicolumn{1}{c|}{0.864}          & 0.593          & \multicolumn{1}{c|}{0.308}          & 0.185          & \multicolumn{1}{c|}{0.552}          & 0.430          \\ \hline
\multirow{2}{*}{NEU}  & + & \multicolumn{1}{c|}{\textbf{0.933}} & \textbf{0.835} & \multicolumn{1}{c|}{0.397}          & 0.230          & \multicolumn{1}{c|}{0.489}          & 0.390          \\ 
                      & - & \multicolumn{1}{c|}{0.833}          & 0.536          & \multicolumn{1}{c|}{0.397}          & \textbf{0.254} & \multicolumn{1}{c|}{\textbf{0.511}} & \textbf{0.415} \\ \hline
\end{tabular}
}
\caption{HR and NDCG results group by 5 personality traits in \textit{Amazon-beauty}, \textit{Amazon-music}, and \textit{Personality2018} datasets. ``+'' represents the NCF+Soft-labeled model (with personality information), and ``-'' represents the NCF+Same model (without personality information).The best performance is in bold.}
\label{tab:analysis}
\vspace{-8mm}
\end{table}

\section{Discussion}

In this work, we make a preliminary attempt to explore how to automatically infer users' personality traits from product reviews and how the inferred traits can benefit the state-of-the-art automated recommendation processes. Although we observe that recommendation performance is indeed boosted by incorporating personality information, we believe there are several limitations. In the following, we discuss these limitations with potential future directions.  

First, we believe capturing personality from the review texts may lead to selective bias. Introverts are less likely to share their thoughts online while extroverts are more likely to share experiences. This results in an imbalanced personality distribution in our collected data. As shown in the analysis in RQ2 (Section~\ref{sec:RQ2}), extroversion is the most common personality trait of users in our datasets. To address this, in future works, we could utilize other context information to infer users' personalities such as a user's purchase history. Such user behaviours can also reflect personality; for example, open people are more likely to follow popular trends which can be reflected in their purchase history. 

Second, we only conduct experiments on a single basic model, NCF, which may loss of generalization. More advanced models graph recommendation models can be used in the future. 
Third, we conduct empirical experiments on whether personality information benefits recommendation. However, more in-depth investigation is necessary on how personality affects recommendation and users' behavior. In the future, we could conduct a user study to find the causal relationship between personality and recommendation. To be specific, we can develop different marketing strategies for users with different personalities. By observing the effects of different strategies on users' behavior, we can gain a better understanding of how personality affects recommendation. 
Fourth, we find that the \textit{openness}, \textit{conscientiousness} and \textit{neuroticism} features do not have a noticeable impact on the recommendation performance. A possible reason is that OCEAN only contains five types of personality, which might be insufficient to provide enough useful signals to recommendations. A possible solution is to use a more fine-grained personality model than OCEAN; \textit{e.g.,} the MBTI personality model which has a richer, 16-facet personality profile. 

Last, the five personalities are encoded independently of each other in our model. But there is a correlation between these personality traits in real life; \textit{e.g.,} a majority of extroverts are also open. In the future, we can make use of the relationship between personalities, perhaps by defining a hierarchical structure of personality traits and employing graph-based neural networks to encode them. 

\section{Conclusion and Future Works}
\label{sec:conclusion}

In this work, we explore a new way of automatically extracting personality information from review texts and applying it to recommendation systems. We first construct two new datasets based on the Amazon dataset in the beauty and music domains and include OCEAN personality scores automatically inferred by the Receptiviti API, a commercial service. We then analyze the accuracy of using texts to obtain personality profiles and output  personality score distributions. To explore the effectiveness of using personality in current recommendation systems, we conduct a few experiments with the standard neural collaborative filtering (NCF) recommendation algorithm and our  variants, finding that incorporating personality information improves recommendation  performance by 3\% to 28\%. In terms of the relationship between personality and domain, we find that \textit{openness}, \textit{extroversion}, and \textit{agreeableness} are helpful in music recommendation, while \textit{conscientiousness} is most helpful in the beauty recommendation. 

In the future, more advanced models graph recommendation models can be used in the experiments. In addition, collecting more information beyond review texts (\textit{e.g.,} purchase history, browsing history) is a potential direction. Moreover, except for the accuracy-based performance, it is possible to improve the fairness by using the OCEAN model~\cite{melchiorre2020personality}. To explore the inner relationship between personality and recommendation systems, doing a user study is also a possible way to further validate the findings.



\section*{Acknowledgement}
We sincerely appreciate Dr. Liangming Pan's efforts in his help in proofreading this work. 


\bibliographystyle{ACM-Reference-Format}
\bibliography{main}
\end{document}